\documentstyle[twocolumn,aps,prb,epsf]{revtex}

\begin{document}
\title{Cluster Spin Glass Distribution Functions in La$_{2-x}$Sr$_x$CuO$_4$
\thanks{This work has been carried out in the framework of the Progetto di
Ricerca Avanzato INFM-SPIS.}}
\author{R.S. Markiewicz$^{(1)}$, F. Cordero$^{(2)}$, A. Paolone$^{(3)}$,
and R. Cantelli$^{(3)}$}
\address{(1) Physics Department and Barnett Institute, \\
Northeastern U., Boston MA 02115; \\
(2) CNR Roma, Istituto di Acustica ''O.M. Corbino'', Roma, Italy; \\
(3) Dept. of Physics, University of Roma ''La Sapienza'' and Unit\`a INFM di 
Roma, Italy}
\maketitle

\begin{abstract}
Signatures of the cluster spin glass have been found in a variety of
experiments, with an effective onset temperature $T_{on}$ that is frequency
dependent. We reanalyze the experimental results and find that they are
characterized by a distribution of activation energies, with a nonzero glass
transition temperature $T_g(x)<T_{on}$. While the distribution of activation
energies is the same, the distribution of weights depends on the process.
Remarkably, the weights are essentially doping independent.
\end{abstract}


\narrowtext

\section{Introduction}

Hole doping in the cuprates is now widely believed to lead to strongly
inhomogeneous phases, with the holes being ejected from the
antiferromagnetic domains onto antiphase boundaries. Ordering of these
charged stripe walls leads to a form of domain phase, which is typically
dynamic. In the very low doping limit, the inhomogeneity is manifested in
the form of transverse spin freezing in the antiferromagnetic phase,\cite{SF}
and, when the N\'{e}el temperature $T_{N}\rightarrow 0$, as a `cluster spin
glass' (CSG).\cite{CSG} Wakimoto, et al. (WUEY)\cite{Waki} have correlated a
number of experimental observations of this CSG in La$_{2-x}$Sr$_{x}$CuO$%
_{4} $ (LSCO), and raise the issue of whether there is a real finite
temperature glass transition, or just a continuous relaxational slowing
down. On the one hand, all of the experiments find the transition at a
different temperature, depending on the time scale of the measurement; on
the other hand, the susceptibility shows striking scaling behavior, both
with temperature and applied field. 

Here, we analyze these data in more detail, combining them with additional
measurements at intermediate frequencies. In all measurements, we find that
the CSG freezing turns on at an onset temperature $T_{on}$ which scales with
the logarithm of the measurement frequency, consistent with a relaxational
form, 
\begin{equation}
I\simeq {\frac{\omega \tau }{1+\omega ^{2}\tau ^{2}}}  \label{eq:1}
\end{equation}
with Vogel-Fulcher\cite{VF} relaxation
\begin{equation}
\tau =\tau _{0}e^{E/(T-T_g)}  \label{eq:2}
\end{equation}
(here $T_{on}$ corresponds to the temperature below which $\tau ^{-1}$
becomes slower than the measuring angular frequency $\omega $). In order to
explain the observed scaling behavior, we find that the CSG must have a
broad distribution of activation energies, with a well defined
characteristic energy scale $E_{m}$, and that different experiments probe
different aspects of this distribution.

\section{$T_g$ and Activation Energy Distribution Functions}

The glass transition temperature $T_{g}$ is typically found by looking for 
deviations from activated behavior: curvature in a plot of $ln(\omega )$ vs. 
$1/T$ (here, $\omega $ is a characteristic frequency of a given
experiment and $T$ is the temperature). The presence of a broad distribution
of activation energies makes it difficult to extract a unique value of $T_{g}
$. For the CSG, there is an additional complicating factor: the
characteristic frequencies are often not well defined. For instance, in the
neutron diffraction experiments,\cite{Waki2} the effective frequency is
taken as the energy resolution 0.25meV: the diffraction peak is elastic, or
static, on that energy scale. On the other hand, the magnetic susceptibility
was measured by a `static' SQUID technique;\cite{Waki} since a typical scan
takes about 10~s,\cite{Waki1} we assume an effective frequency of $\sim 1/20$%
~Hz. Clearly, improved determination of these frequencies will lead to
better estimates for $T_{g}$ -- particularly at the lowest frequencies.
Nevertheless, the variation with frequency is so striking that the
qualitative features should be unchanged by these refinements.

The simplest distribution comes from measuring the magnetic susceptibility $%
\chi _{0}(T)$. WUEY found that the in-plane $\chi _{0}$ could be fit to a
simple Curie law with small Curie constant for temperatures between 10~K and
70~K, while at lower temperatures the Curie constant became temperature
dependent as clusters began to freeze out. To analyze these results, we
assume that the susceptibility has a relaxational component 
\begin{equation}
\chi _{0}(T)\equiv \chi _{C}(T)(1-q)=\int dE\,w_{susc}(E){\frac{\chi _{C}}{%
1+\omega ^{2}\tau (E)^{2}}},  \label{eq:3}
\end{equation}
with $\chi _{C}$ the Curie susceptibility and $w_{susc}(E)$ the
susceptibility distribution function. This distribution function has a
simple interpretation in the standard model of 
\begin{figure}[tbp]
\leavevmode
   \epsfxsize=0.33\textwidth\epsfbox{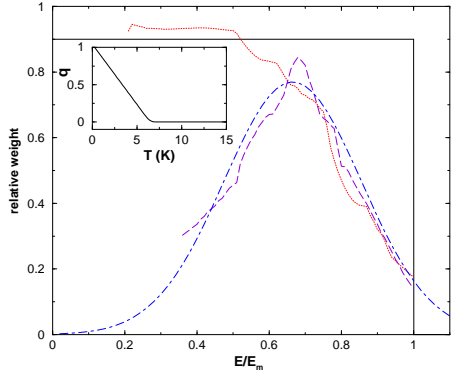}
\vskip0.5cm
\caption{Distributions of excitation energies $w_{susc}$ (solid line), $%
w_{anelast}$ (dotted line), and $w_{NQR}$ (dashed line); dot-dashed line =
Gaussian approximation to $w_{NQR}$. Insert: resulting $q(T)$, Eq.\ref{eq:3}%
. }
\label{fig:1}
\end{figure}
\par\noindent
the susceptibility. It is
assumed\cite{Good} that there is a distribution of clusters, with their
spins locked in an antiferromagnetic pattern, so that each cluster with an
even (odd) number of spins has net spin zero (1/2). The observed Curie law
is attributed to the response of the odd-spin clusters, thereby naturally
explaining the small Curie constant. Since the Curie law is obeyed between
10-70~K, the clusters must have a well defined size distribution, which
begins to freeze out at lower temperatures; $w_{susc}(E)$ is then the
distribution of the {\it number} of clusters with a given activation energy,
independent of the cluster size. If this distribution is normalized ($\int
dE\,w_{susc}(E)=1$), then 
\begin{equation}
q=\int dE\,w_{susc}(E){\frac{\omega ^{2}\tau (E)^{2}}{1+\omega ^{2}\tau
(E)^{2}}}.  \label{eq:4}
\end{equation}
The strong temperature dependence of $\tau $, Eq.\ref{eq:2}, means that Eq. 
\ref{eq:4} can be simplified. For each temperature, there is a corresponding
activation energy for which $\omega \tau (E(T))=1$, namely $%
E(T)=-T\,ln(\omega \tau _{0})$. Then the effective Edwards-Anderson\cite{EA}
order parameter $q$ may be approximated 
\begin{equation}
q\simeq \int_{E(T)}^{E_{m}}dE\,w_{susc}(E).  \label{eq:5}
\end{equation}
The observed\cite{Waki} scaling $q\sim (T_{on}-T)^{\beta }$ with $\beta \sim
1$ then suggests that $w_{susc}(E)=$ const for $E\le E_{m}$, with $%
E_{m}=E(T_{on})$. Thus the appearance of $T_{on}$ is interpreted as the
presence of a sharp cutoff $E_{m}$, 
\begin{equation}
T_{on}=-E_{m}/ln(\omega \tau _{0}).  \label{eq:6}
\end{equation}
The (flat) distribution of $w_{susc}$ is shown in Figure~\ref{fig:1}; the
insert to the figure shows that the above analysis is well satisfied by
direct numerical integration of Eq.~\ref{eq:3}. We note that Eq.~\ref{eq:3}
has been applied\cite{FH} for describing susceptibility data in the spin
glass state ($T<T_{g}$); however, the above analysis depends only on the
form of the distribution, and holds even if $T_{g}=0$. 

The above analysis can be extended to other experimental measurements. There
are two changes. First, the susceptibility is proportional to the product of
a coupling and a distribution function, $\chi _{C}(T)\,w_{susc}(E)$; for
other properties, this can be written $A_{i}(T)\,w_{i}(E)$, where $A_{i}$
represents the intrinsic temperature dependence of the process. In these
calculations, we assume that $A_{anelast}\sim 1/T$ (Ref. \onlinecite{NB})
while $A_{NQR}\sim const$. Secondly, since most of the other probes are
sampling a dissipation, there is an extra factor of $\omega \tau $ in the
integrand, as in Eq.\ref{eq:1}. This factor means that the integrand at any $%
T$ will be strongly peaked at $E(T)$. 
\begin{figure}[tbp]
\leavevmode
   \epsfxsize=0.40\textwidth\epsfbox{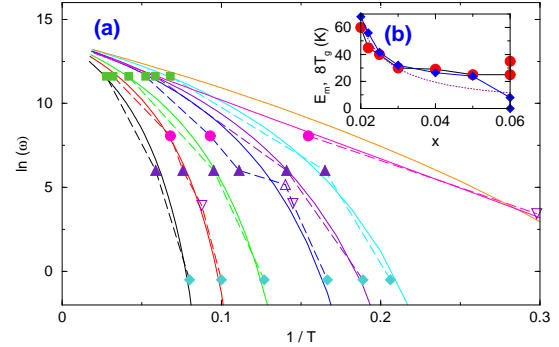}
\vskip0.5cm
\caption{(a) Scaling of effective transition temperature (corresponding to
largest activation energy) with frequency. The symbols represent the
measurement techniques: squares = neutron diffraction; circles = NQR; filled
triangles = $\mu$SR; open, inverted triangles = anelastic relaxation (1st
mode -- open triangle = 5th mode); and diamonds = susceptibility.
The different sets connected by dashed lines refer to individual doping
densities, from left to right: $x$ = 0.02, 0.022, 0.025, 0.03, 0.04, 0.05,
and 0.06. Solid lines = fits. Insert (b): corresponding $T_g$ (diamonds) and 
$E_m(x)/8$ (circles), with dotted line = fit. $T_g$ is multiplied by 8 for
comparison with $E_m$.}
\label{fig:2}
\end{figure}

Figure~\ref{fig:2}a summarizes the doping dependence of $T_{on}$ vs.
measurement frequency. The data are, from highest frequency: squares =
neutron scattering determination of the onset of (diagonal) stripe order\cite
{Waki2} ($\omega $ = 0.25~meV = $4\times 10^{11}$s$^{-1}$); circles = NQR
measurements\cite{Cord} ($\omega /2\pi $ = 19~MHz); filled triangles = $\mu $%
SR\cite{Nied} ($\omega $ = 10$^{6}$~s$^{{-1}}$); open triangles = anelastic
relaxation\cite{Cord} ($\omega /2\pi $ $\simeq $ 2~kHz); and diamonds =
susceptibility\cite{Waki} ($\omega /2\pi $ $\simeq $ $1/20$~Hz). Solid lines
are fits assuming a finite glass transition temperature $T_{g}$; Figure~\ref
{fig:2}b shows the parameters $T_{g}\left( E_{m}\right) $ assumed in the
fits. [In order to plot Fig.~\ref{fig:2}, it was sometimes necessary to use
the smoothed curves of Ref.~\onlinecite{Waki} rather than the actual data;
this would only be a problem in the doping range $x\le 0.025$, where some
curves are extrapolated. Also, two fits are shown for the $x=0.06$ data: the
straight line corresponding to the larger $E_{m}$ and smaller $T_{g}$ ($=0$%
).] While there is considerable scatter in the data, it appears that $\tau
_{0}$ is essentially doping independent, while $E_{m}$ decreases with
doping, Fig.~\ref{fig:2}b. For most dopings, the approximate relation $%
E_{m}=8k_{\text{B}}T_{g}$ is satisfied. The value of $\tau _{0}^{-1}=5\times
10^{13}~$s$^{-1}=33~$meV is close to the value 22~meV estimated by Julien
\thinspace  {\it et al.} \cite{Jul} from the peak in the NQR $T_{1}^{-1}$.
The dotted line in Fig.~\ref{fig:2}b is a fit to the form $%
E_{m}=E_{0}/(x-x_{0})$, with $E_{0}=0.564~$K, $x_{0}=0.011$. The general
trend $E_{m}\sim 1/x$ has been noted earlier.\cite{BCRR}

We have included in Fig.~\ref{fig:2} some anelastic and NQR data for a
sample with $x\sim 0.02$. Uncertainty in the precise doping makes it unclear
whether the actual doping is $x\simeq 0.017$, with spin freezing transition $%
T_{f}\sim 13.9$~K or $x\gtrsim 0.02$, with $T_{on}\sim 10$~K. We find that
the assumption of an effective density $x_{eff}=0.021$ yields remarkable
agreement for both $T_{on}(\omega )$, but also for the weight distributions
for both anelastic relaxation and NQR\cite{JCB}, as illustrated in Figs.~\ref
{fig:4} and \ref{fig:5} below. Indeed, results by Curro, et al.\cite{Cur}
show that La NMR weight distributions are very similar in the spin freezing
and CSG regimes.

\begin{figure}[tbp]
\leavevmode
   \epsfxsize=0.40\textwidth\epsfbox{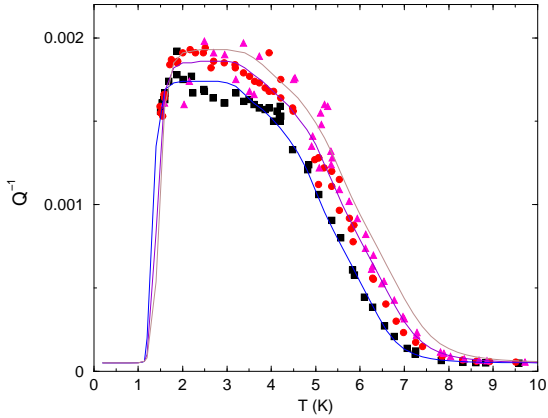}
\vskip0.5cm
\caption{Elastic energy loss coefficient, $Q^{-1}(T)$, for a $x =$ 0.03
sample, by exciting the lowest three (odd) flexural modes: squares = 1st ($%
\omega /2\pi = 1.7$~kHz), circles = 3rd (9.3~kHz), triangles = 5th
(23~kHz). Solid lines = fit to the first mode, with distribution $%
w_{anelast}$ shown in Fig. 1, scaled by $\omega$ for the higher modes.}
\label{fig:3}
\end{figure}

Figure~\ref{fig:3} illustrates the fit of the elastic energy loss
coefficient $Q^{-1}$ for the three lowest odd modes of an LSCO sample at $%
x=0.03$, while the corresponding distribution function is shown in Figure~%
\ref{fig:1}. Approximately the same distribution works for the $x\sim 0.02$
and the $x=0.06$ samples (Fig.~\ref{fig:4}). Figure~\ref{fig:5} shows the
corresponding fits for the NQR relaxation rate; the distribution function
(shown scaled in Fig.~\ref{fig:1}) was fit for the $x\sim 0.02$ sample, then
scaled for the others. Note that for the anelastic response the fit is worse
for the higher harmonics, and for $x\sim 0.02$ anelastic data there is an
additional, lower-temperature process with a very different doping
dependence.\cite{Cor2} Also there appears to be a weak extra feature in the
NQR ($\sim 0.02$) data near 20~K. It should be cautioned that the inversion
of the data to obtain the distributions, Fig.~\ref{fig:1}, is not unique:
the lorentzians are sufficiently broad that several distributions could lead
to the same final spectra. The distributions chosen in Fig.~\ref{fig:1} were
chosen to be relatively smooth. [Note that the low-energy cutoffs are
exaggerated, since data at sufficiently low temperatures are lacking.] 
\begin{figure}[tbp]
\leavevmode
   \epsfxsize=0.40\textwidth\epsfbox{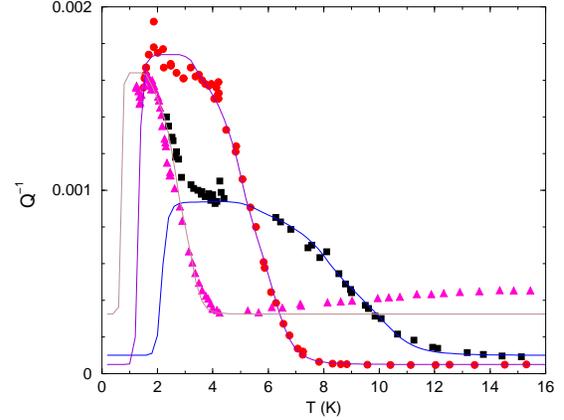}
\vskip0.5cm
\caption{Fit to the elastic energy loss coefficient, $Q^{-1}(T)$, for
samples of several dopings, $x$ $\sim$ 0.02 (squares), 0.03 (circles), and
0.06 (triangles).\protect\cite{JCB} All fits use the same distribution of $%
w_{anelast}$.}
\label{fig:4}
\end{figure}

All three distributions in Fig.~\ref{fig:1} appear to be universal, in that
the same distribution holds over a wide doping range. However, different
properties appear to be characterized by different distribution functions.
We suggest that the different experiments may sample different attributes of
the clusters. Thus, as discussed above, the susceptibility is proportional
to the {\it number} of clusters with a particular activation energy. Since
anelastic relaxation responds to the change in elastic energy when one type
of domain grows at the expense of another domain with less favorable
orientation, $w_{ anelast}$ should be sensitive to the domain walls, namely
the {\it perimeters} of the domains; similarly $w_{NQR}(E)$ should measure
the total number of spins, or the total {\it areas} of all the domains with
a given activation energy $E$.

The nearly constant $Q^{-1}$ at low $T$ implies a nearly constant
distribution $w_{anelast}$ at low $E$, Fig.~\ref{fig:1}; this can be
understood as follows: since 
\begin{eqnarray}
Q^{-1}=\int dE{\frac{w_{anelast}(E)}{T}}{\frac{\omega\tau}{1+\omega^2
\tau(E)^2}}  \nonumber \\
\simeq\int dE{\frac{w_{anelast}(E)}{T}}\delta (\omega\tau -1)  \nonumber \\
={\frac{w_{anelast}(E(T))}{T|{\frac{\partial\omega\tau}{\partial E}}%
|_{\omega\tau=1}}} =w_{anelast}(E(T)),  \label{eq:7}
\end{eqnarray}
so when $Q^{-1}$ is roughly constant, so is $w_{anelast}(E(T))$. 
\begin{figure}[tbp]
\leavevmode
   \epsfxsize=0.40\textwidth\epsfbox{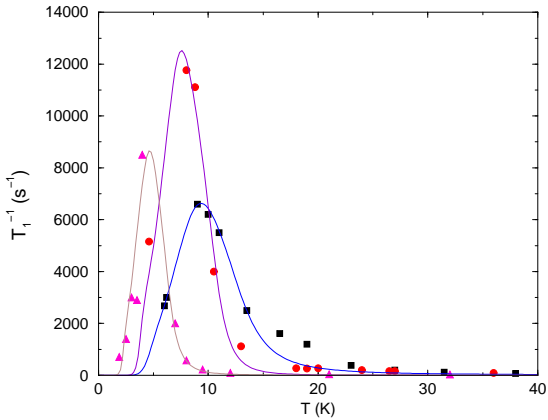}
\vskip0.5cm
\caption{Fit to NQR data $T_1^{-1}$ for several dopings, $x$ $\sim$ 0.02
(squares), 0.03 (circles), and 0.06 (triangles).  The data are from Refs.
\protect\onlinecite{Cord} and \protect\onlinecite{Jul}.
All fits use the same distribution of $w_{NQR}$.}
\label{fig:5}
\end{figure}

\section{Discussion}

The distributions, Fig.~\ref{fig:1}, and the spin glass transition
temperatures $T_g$ are the principle results of this work. As discussed
above, $w_{susc}$ measures the cluster number distribution, $w_{anelast}$
the perimeter distribution, and $w_{NQR}$ the area distribution. Figure~\ref
{fig:1} suggests that the perimeters and areas follow the same distribution
at high energies, but that the area distribution starts to fall off sooner.

An important finding is the {\it universality} of the distributions: the
same distributions appear to hold over a wide doping range. Earlier La NMR
measurements found a similar broad distribution (fit to a Gaussian)\cite{Suh}%
, and found a universal weight distribution over an even wider doping range,
including spin-freezing, near-optimal superconducting, and Li-doped samples.%
\cite{Cur} Indeed, the present NQR distribution can be fit to a Gaussian,
dashed line in Fig.~\ref{fig:1}, 
\begin{equation}
w={\frac{w_{0}}{\sqrt{2\pi }z_{A}}}e^{{\frac{(z-z_{B})^{2}}{2z_{A}^{2}}}},
\label{eq:66}
\end{equation}
with $z=E/E_{m}$, $w_{0}=0.367$, $z_{A}=0.190$, and $z_{B}=0.665$.

It is possible to use the distribution functions to understand how the spin
slowing down evolves with temperature, since (at any given frequency) the
processes with the highest values of $E(T)$ will freeze out first. Thus the
distribution $w(E)$ can be alternatively considered as a distribution $w(T)$
of the number of processes freezing out at temperature $T$, with $T\sim E/k_{%
\text{B}}$. From the neutron scattering, at $T_{on}(\omega )$ diagonal
stripes start to order, with growing coherence length, but the coherence
length cuts off at a fairly small value without diverging. At the same time
the number of free clusters is decreasing (from $w_{susc}$). There is a
large change in the number of clusters ($w_{susc}$), but only a small
decrease in the total number of fluctuating spins ($w_{area}=w_{NQR}$), so
presumably the smallest, most strongly pinned clusters (largest $E(T)$) are
freezing out. This interpretation that the clusters are small is
strengthened by the fact that they make approximately the same contribution
to $w_{perim}=w_{anelast}$ and $w_{area}=w_{NQR}$ (i.e., all spins are on
the surface of the cluster). The fact that $w_{area}$ is finite suggests
that the spins are freezing out rather than clustering. At some point
(perhaps when the correlation length stops increasing) the contribution to $%
w_{area}$ drops sharply, but the cluster walls can still move. Note that
even though $w_{area}$ is shrinking, there are still many freely moving
clusters ($w_{susc}$ remains constant), suggesting that something like a
`percolating backbone' has frozen out, and at lower temperatures the
backbone grows by capturing additional clusters.

The observed curvature in the plot of $ln(\omega )$ vs. $1/T$ strongly
suggests the existence of a well-defined finite temperature CSG transition
at fixed $T_{g}$. Additional measurements, especially with well-defined low
frequencies, would help pin down the value of $T_{g}(x)$, but should not
change the overall picture. The interpretation is complicated by the
presence of a broad distribution of activation energies, which lead to
slowing down over a wide frequency range. In particular, we have shown that
an apparent scaling of the order parameter $q$ can arise from a particular
form of $w(E)$, with sharp upper cutoff $E_{m}$. However, scaling behavior
is also found in an applied magnetic field,\cite{Waki} and this is harder to
explain. The doping dependence of $E_{m}$ (insert, Fig.~\ref{fig:2}) is very
suggestive of pinning effects. Note that $E_{m}$ appears to diverge as $%
x\rightarrow x_{0}\sim 0.01$, but this is cut off by the N\'{e}el transition
near $x=0.02$. We suggest that the observed behavior is characteristic of
stripe freezing in the presence of defect pinning. In the absence of
pinning, there could be a long range stripe ordering phase transition, but
only at $T\rightarrow 0$ due to fluctuations of the walls in two dimensions.
In this case, the correlation length would diverge as $T\rightarrow 0$.
Impurities will tend to pin the charge stripes, but the pinning will vary in
space, depending on the distribution of the impurities, and hence leading to
a wide distribution of activation energies. Certainly, the large attempt
frequency $\tau _{0}\sim 2\times 10^{-14}~$s (Fig.~\ref{fig:2}) is
suggestive of an electronic process, $\hbar \tau _{0}^{-1}\sim J/4$, with $J$
the exchange constant.

In attempting to refine the experimental determination of $T_{g}(x)$, it
must be noted that $T_{on}$ may be a strong function of the sensitivity of
the probe to fluctuations. For instance, whereas the La NQR shows a $T_{on}$
in the range of $\sim 10$-20~K, Cu NMR is wiped out at a much higher
temperature, $\sim 50$~K. It has been suggested\cite{Cur,Jul} that this
wipeout is caused by the same CSG fluctuations, which are in the CuO$_{2}$
plane, and hence have a much greater effect on Cu than on La (for an
alternative interpretation, see Ref. \onlinecite{Hunt}). 
A remaining puzzle is that the wipeout should start to
recover below $T_{g}$,\cite{ChSl} whereas no recovery is found for the Cu
signal down to 350~mK.\cite{Ima}

\section{Conclusion}

The signatures of spin freezing in La$_{2-x}$Sr$_{x}$CuO$_{4}$ from neutron
spectroscopy, NQR relaxation, anelastic relaxation and magnetic
susceptibility (spanning $\sim 13$ orders of magnitude of frequency)
have been reconsidered in terms of a distribution $w\left( E\right) $ of
activation energies for the spin dynamics. Each experiment probes a
different aspect of the spin clusters, like the total number of spins in a
cluster, the spins along its perimeter or the number of clusters with an
unpaired spin, and therefore requires a different weight function for the
distribution of activation energies. It is found, however, that for each process
the same $w\left( E/E_{m}\left( x\right) \right) $ can explain all the data, 
independent of doping, with the
same $E_{m}\left( x\right) $ for all the experiments. In addition, $%
E_{m}\left( x\right) $ scales linearly with the temperature $T_{g}\left(
x\right) $ at which the effective spin fluctuation rate deduced from all the
experiments freezes toward the cluster-spin glass state.

{\bf Acknowledgments:}
The authors thank Prof. A. Rigamonti for useful advice. One of
the authors (R.S. Markiewicz) wishes to thank the University of Rome ``La
Sapienza" for its hospitality.


\begin{references}
\bibitem{SF}  F.C. Chou, F. Borsa, J.H. Cho, D.C. Johnston, A. Lascialfari,
D.R. Torgeson, and J. Ziolo, Phys. Rev. Lett. {\bf 71}, 2323 (1993).

\bibitem{CSG}  J.H. Cho, F. Borsa, D.C. Johnston, and D.R. Torgeson, Phys.
Rev. B{\bf 46}, 3179 (1992).

\bibitem{Waki}  S. Wakimoto, S. Ueki, Y. Endoh, and K. Yamada, Phys. Rev. B%
{\bf 62}, 3547 (2000).

\bibitem{VF} H. Vogel, Phys. Z. {\bf 22}, 645 (1921); G.S. Fulcher, J. Am. Chem.
Soc. {\bf 8}, 339 (1925).

\bibitem{Waki2}  S. Wakimoto, G. Shirane, Y. Endoh, K. Hirota, S. Ueki, K.
Yamada, R.J. Birgeneau, M.A. Kastner, Y.S. Lee, P.M. Gehring, and S.H. Lee,
Phys. Rev. B{\bf 60}, 769 (1999).

\bibitem{Waki1}  S. Wakimoto, personal communication.

\bibitem{Good}  R.J. Gooding, N.M. Salem, R.J. Birgeneau, and F.C. Chou,
Phys. Rev. B{\bf 55}, 6360 (1997).

\bibitem{EA}  S.F. Edwards and P.W. Anderson, J. Phys. F{\bf 5}, 965 (1975).

\bibitem{FH}  K.H. Fischer and J.A. Hertz, ``Spin Glasses'' (Cambridge,
Cambridge University Press, 1991).

\bibitem{NB}  A.S. Nowick and B.S. Berry {\it Anelastic Relaxation in
Crystalline Solids} (Academic, 1973).

\bibitem{Cord}  A. Campana, M. Corti, A. Rigamonti, R. Cantelli, and F.
Cordero, cond-mat/0005326, to be published, Europ. Phys. J.

\bibitem{Nied}  Ch. Niedermeyer, C. Bernhard, T. Blasius, A. Golnik, A.
Moodenbaugh, and J.I. Budnick, Phys. Rev. Lett, {\bf 80}, 3843 (1998).

\bibitem{Jul}  M.-H. Julien, A. Campana, A. Rigamonti, P. Carretta, F.
Borsa, P. Kuhns, A.P. Reyes, W.G. Moulton, M. Horvati\'{c}, C. Berthier, A.
Vietkin, and A. Revcolevschi, cond-mat/0010362.

\bibitem{BCRR}  F. Borsa, M. Corti, T. Rega, and A. Rigamonti, Nuovo Cimento 
{\bf 11D}, 1785 (1989).

\bibitem{JCB}  M.-H. Julien, P. Carretta, anf F. Borsa, cond-mat/9909351.

\bibitem{Cur}  N.J. Curro, P.C. Hammel, B.J. Suh, M. H\"{u}cker, B.
B\"{u}chner, U. Ammerahl, and A. Revcolevschi, Phys. Rev. Lett. {\bf 85},
642 (2000).

\bibitem{Cor2}  F. Cordero, R. Cantelli and M. Ferretti, Phys. Rev. B {\bf 61%
}, 9775 (2000).

\bibitem{Suh}  B.J. Suh, P.C. Hammel, M. H\"{u}cker, B. B\"{u}chner, U.
Ammerahl, and A. Revcolevschi, Phys. Rev. B{\bf 61}, 9265 (2000).

\bibitem{Hunt}  A.W. Hunt, P.M. Singer, K.R. Thurber, and T. Imai, Phys.
Rev. Lett. {\bf 82}, 4300 (1999); P.M. Singer, A.W. Hunt, A.F
Cederstr\"{o}m, and T. Imai, Phys. Rev. B{\bf 60}, 15345 (1999).

\bibitem{ChSl}  M.C. Chen and C.P. Slichter, Phys. Rev. B{\bf 27}, 278
(1983).

\bibitem{Ima}  A.W. Hunt, P.M. Singer, A.F. Cederstrom, and T. Imai, 
cond-mat/0011380. 
\end{references}
\end{document}